\documentclass[%
superscriptaddress,
    preprint,
 aps,
prl,
]{revtex4-2}

\usepackage[T1]{fontenc}
\usepackage{lmodern}
\usepackage{graphicx}
\usepackage{dcolumn}
\usepackage{bm}
\usepackage{comment}
\usepackage{xcolor}

\usepackage{siunitx}

\begin{document}

\title{Interfacial Orbital Rashba Edelstein Effect Identified in CoO/Cu/Pt Trilayers}

\author{C. Schmitt}
\thanks{equal contribution}
\email{Christin.Schmitt@Uni-Mainz.de}
\affiliation{Institute of Physics, Johannes Gutenberg University Mainz, 55099 Mainz, Germany}
\author{S. Krishnia}
\thanks{equal contribution}
\email{Krishnia@Uni-Mainz.de}
\affiliation{Institute of Physics, Johannes Gutenberg University Mainz, 55099 Mainz, Germany}
\author{E. Galíndez-Ruales}
\affiliation{Institute of Physics, Johannes Gutenberg University Mainz, 55099 Mainz, Germany}
\author{H. Meer}
\affiliation{Institute of Physics, Johannes Gutenberg University Mainz, 55099 Mainz, Germany}

 \author{T. Kikkawa}
\affiliation{Department of Applied Physics, The University of Tokyo, Tokyo 113-8656, Japan}
\affiliation{Advanced Science Research Center, Japan Atomic Energy Agency, Tokai 319-1195, Japan}
\author{T. Denneulin}
 \affiliation{Ernst Ruska-Centre for Microscopy and Spectroscopy with Electrons, Forschungszentrum J\"ulich, 52425 J\"ulich, Germany.}
 \author{A. Kov\'acs}
 \affiliation{Ernst Ruska-Centre for Microscopy and Spectroscopy with Electrons, Forschungszentrum J\"ulich, 52425 J\"ulich, Germany.}
\author{O. Kuschel}
 \affiliation{Institute of Physics, Johannes Gutenberg University Mainz, 55099 Mainz, Germany}
  \affiliation{Department of Physics and Center of Physics and Chemistry of New
Materials, Osnabrück University, 49076 Osnabrück, Germany}
 \author{R. E. Dunin-Borkowski}
 \affiliation{Ernst Ruska-Centre for Microscopy and Spectroscopy with Electrons, Forschungszentrum J\"ulich, 52425 J\"ulich, Germany.}
\author{E. Saitoh}
\affiliation{Department of Applied Physics, The University of Tokyo, Tokyo 113-8656, Japan}
\affiliation{RIKEN Center for Emergent Matter Science, Wako 0351-0198, Japan}
\affiliation{WPI-Advanced Institute for Materials Research, Tohoku University, Sendai 980-8577, Japan}
 \affiliation{The Institute of AI and Beyond, The University of Tokyo, Tokyo 113-8656, Japan}
\author{T. Kuschel}
 \affiliation{Institute of Physics, Johannes Gutenberg University Mainz, 55099 Mainz, Germany}
\author{Y. Mokrousov}
 \affiliation{Institute of Physics, Johannes Gutenberg University Mainz, 55099 Mainz, Germany}
  \affiliation{Peter Gr\"unberg Institut and Institute for Advanced Simulation, Forschungszentrum J\"ulich and JARA, 52425 J\"ulich, Germany}
\author{M. Kläui}
\email{Klaeui@Uni-Mainz.de}
 \affiliation{Institute of Physics, Johannes Gutenberg University Mainz, 55099 Mainz, Germany}
 \affiliation{Center for Quantum Spintronics, Norwegian University of Science and Technology, 7491 Trondheim, Norway}

\date{October 25, 2026}

\begin{abstract}

We investigate the generation of orbital angular momentum currents at engineered Cu–oxide interfaces using CoO/Cu/Pt thin-film trilayers. Angular-dependent magnetoresistance measurements on samples with varying Cu interlayer thickness reveal systematic changes in both longitudinal and transverse resistance components, including a sign inversion of the signal. By systematically varying the Cu  interlayer thicknesses, we disentangle the spin and orbital contributions to the magnetoresistive response. Our results show that large orbital currents originate from the interfacial orbital Rashba–Edelstein effect due to the presence of the Cu–O bonds at the CoO/Cu interface. Unlike previously studied CoO/Cu bilayers that rely on uncontrolled natural oxidation, our trilayer geometry enables reproducible and stable orbital-current generation at ambient conditions. Our results identify Cu–oxide interfaces as efficient sources of orbital angular momentum and clarify the microscopic mechanism underlying orbital current generation. Consequently, this work establishes a well-defined materials platform for all-orbitronic devices that does not require relativistic spin–orbit coupling or orbital-spin interconversion processes.
\end{abstract}

\maketitle

\newpage

\newpage
\section{\label{sec:intro}Introduction\protect}

Controlling and detecting magnetic order through angular momentum transport has been a cornerstone of spintronics technologies~\cite{jungwirth2016antiferromagnetic, Manchon2019}. While spin currents have long been the dominant mechanism for such control, their generation and efficiency are fundamentally limited by the relativistic spin-orbit coupling (SOC), which is inherently weak due to its relativistic nature~\cite{Manchon2019}. Consequently, the choice of suitable materials for spin current generation is restricted, and the scalability and energy efficiency of emerging applications are compromised.
A promising alternative to spin-based approaches is the use of orbital angular momentum (OAM)~\cite{go2021orbitronics,DongwookPRL2018,Lee2024,fukami2025challenges}. OAM current can be generated efficiently from charge currents due to direct coupling between the crystal momentum of charge carriers and their orbital degrees of freedom~\cite{DongwookPRL2018,salemi2022first,krishnia2023,Choi2023, Kawakami2023, idrobo2024}. Consequently, orbital currents can be generated via the orbital Hall effect (OHE) in the bulk~\cite{Tanaka2008, Kontani2009,DongwookPRL2018} and the orbital Rashba-Edelstein effect (OREE) at interfaces~\cite{Johansson2021,Nikolaev2024,Ding2022}, both of which are effective even in materials with negligible SOC~\cite{DongwookPRL2018,salemi2022first,krishnia2023,Choi2023, Kawakami2023, idrobo2024}. Recent theoretical and experimental studies have demonstrated that orbital currents can be produced with significantly higher efficiency than spin currents, even in light and abundant materials~\cite{salemi2022first,Shilei2020}. This has positioned orbitronics, the use of OAM for information transport and manipulation, as a promising avenue for low-power, high-efficiency magnetic devices.
A key challenge, however, has been that conventional 3$d$ ferromagnets, in which magnetism arises predominantly from spin angular momentum~\cite{CTChenPRL1995,krishnia2024interfacial}, interact only weakly with orbital currents, as the orbital moments are largely quenched. Consequently, most demonstrations of orbitronic functionality have required a conversion step in which orbital currents are transformed into spin currents using high-SOC materials~\cite{Shilei2020,ding2024orbital}. This conversion re-introduces the efficiency constraints associated with SOC, limiting the practical advantage of large orbital currents.

A recent breakthrough demonstrated that orbital currents can couple directly to OAM-dominated magnetic materials, bypassing the need for orbital to spin conversion~\cite{Schmitt2026}. In insulating antiferromagnetic $\mathrm{CoO/Cu}^*$ heterostructures, orbital currents generated in the surface-oxidized Cu (Cu*) layer interact directly with the unquenched orbital angular momentum of CoO, thereby enabling efficient magnetization manipulation without relying on weak relativistic SOC effects~\cite{Schmitt2026}. This development marks a significant step toward realizing high-efficiency orbitronic devices. However, the role of CoO/Cu interface in generating the larger magnetoresistance (MR) signal and torques is currently an open question. Although calculations and experiments suggest that the OREE at the Cu/CuO$_x$ interface dominates the OAM generation~\cite{go2021orbital,Krishnia2024,Bony2025,Shilei2020}, additional contributions from the interface of Cu with an oxide cannot be excluded. In our previous study, the orbital current could be generated either at the oxidized Cu/CuO$_x$ surface or also at the CoO/Cu interface~\cite{Schmitt2026}. Elucidating the underlying OAM current generation mechanism is therefore essential for optimizing material stacks and ensuring the long-term stability of orbitronic devices. This is particularly important given that uncontrolled surface oxidation of the Cu layer can lead to unstable operating conditions due to time- and environment-dependent variations in oxidation. In this work, we address this open question by investigating the dependence of MR on Cu thickness in CoO/Cu/Pt trilayers. By varying the Cu thickness while keeping Pt fixed, we control the relative distribution of the total charge current flowing in the two metals, thereby tuning the relative weights of the orbital- and spin-related effects. Our results allow us to identify the roles played by the interfaces for spin and orbital current generation. By tuning the layers, we can engineer stable and robust devices with controlled interfaces, enabling the development of more efficient and stable orbitronic applications.

\section{\label{sec:results}Results\protect}

We deposit MgO(001)//CoO($\SI{5}{nm}$)/Cu($t_{\mathrm{Cu}}$)/Pt($\SI{2}{nm}$) heterostructures by reactive magnetron sputtering, with the thickness $t_{\mathrm{Cu}}$ of the Cu layer varying between $\SI{0}{nm}$ and $\SI{25}{nm}$ (details, see supplementary information-1). The Cu and Pt layers are deposited on the CoO in-situ without breaking the vacuum. Given that the interface quality crucially influences the signal quality in spin Hall magnetoresistance (SMR) measurements~\cite{putter2017impact, althammer2019role, fontcuberta2019role, catalano2022spin}, we first characterize the structural and chemical quality of the interfaces
using X-ray reflectometry (XRR) and scanning transmission electron
microscopy (STEM) using high-angle annular dark-field (HAADF) imaging. This step ensures that all samples are comparable in terms of interfacial quality and layer continuity prior to transport characterization.

\begin{figure*}
 \includegraphics[width=15cm]{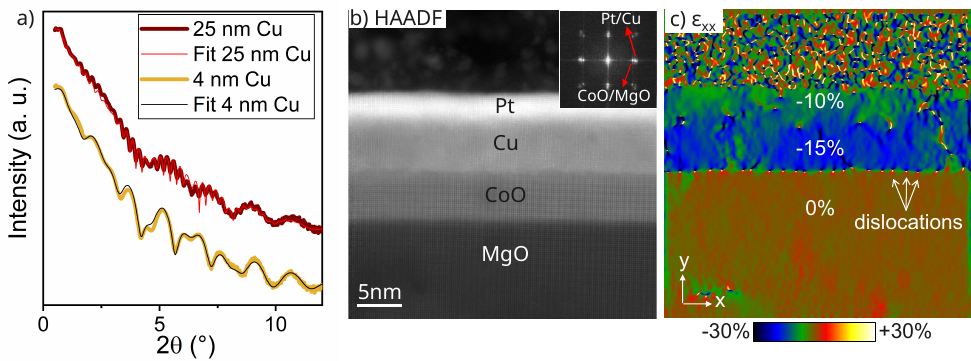}
 \caption{(a) XRR data for two CoO/Cu/Pt films with $\SI{4}{nm}$ and \SI{25}{nm} Cu layer thickness showing flat films with well-defined interfaces. (b) HAADF STEM image and corresponding FFT in the inset. (c) Horizontal  deformation $\varepsilon_{xx}$   map calculated using geometric phase analysis. The deformation is calculated with respect to the lattice spacing of the MgO substrate following $\varepsilon =(d-d_{\mathrm{MgO}})/d_\mathrm{MgO}$ where $d$ is the local lattice spacing.}
 \label{fig:one}
\end{figure*}

Figure \ref{fig:one}(a) shows the XRR data for two representative samples with Cu interlayer thicknesses of $\SI{4}{nm}$ and $\SI{25}{nm}$. Both samples exhibit smooth, well-defined
interfaces, with a surface/interface roughness of the CoO and Cu layers below $\SI{0.4}{nm}$. This high interfacial quality minimizes defect- and boundary-induced scattering, enabling robust spin- and orbital-current transfer and ensuring sample-to-sample comparability. Furthermore, the $\SI{2}{nm}$ thick Pt layer exhibits a maximum roughness of $\SI{0.3}{nm}$, confirming continuous film growth that effectively passivates the Cu interlayer against oxidation from air. To assess the epitaxial quality as well as the chemical and structural characteristics of the CoO/Cu interface, we analyze a HAADF STEM  image of MgO//CoO/Cu/Pt heterostructure shown in Fig. \ref{fig:one}(b). The HAADF-STEM image reveals coherent epitaxial growth of the CoO layer on the MgO substrate, evidencing its high crystalline quality. The corresponding fast Fourier transform (FFT), shown in the inset of  Fig. \ref{fig:one}(b), further indicates that the Cu layer maintains an epitaxial relationship with the underlying CoO, confirming the formation of a well-defined interface that supports subsequent epitaxial growth. 
In addition, the map displayed in  Fig. \ref{fig:one}(c), derived from geometrical phase analysis (GPA) of the image in  Fig. \ref{fig:one}(b) and referenced to the MgO substrate, reveals the spatial variation of the in-plane lattice spacing across the stack. At the CoO/Cu interface, periodic lattice distortions with a characteristic spacing of approximately 1.3 nm are observed, indicating that a regular array of misfit dislocations accommodates the interfacial strain while preserving epitaxial growth of Cu.

\begin{figure*}
 \includegraphics[width=15cm]{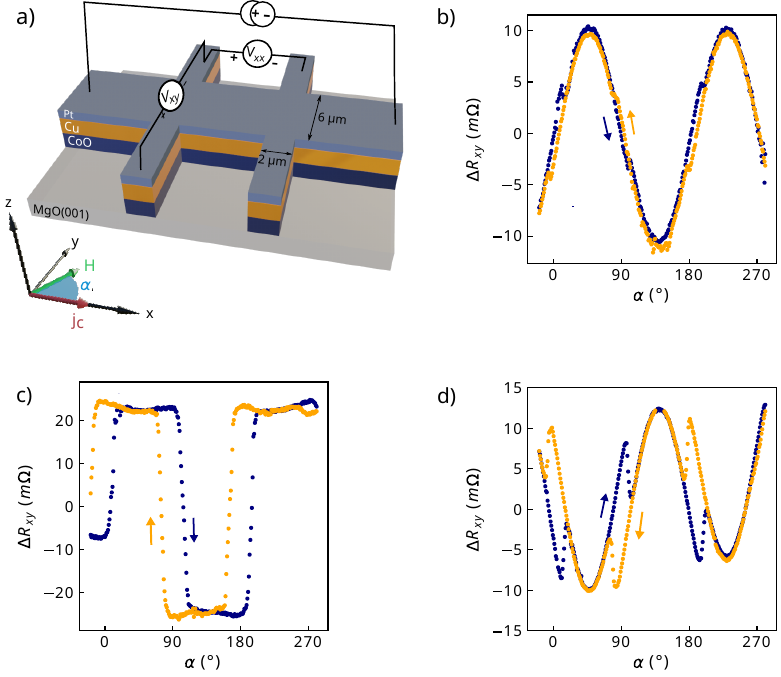}
 \caption{ (a) Illustration of the MR measurement scheme with electrical contacts in the CoO/Cu/Pt structure, including the definition of the angle sweep plane $\alpha$. Transverse $R_{xy}$ MR measurements for (b) CoO(5)/Pt(2), (c) CoO(5)/Cu(3)/Pt(2) and (d) CoO(5)/Cu(6)/Pt(2), respectively.}
 \label{fig:two}
\end{figure*}

For the electrical characterization, Hall bar devices (width $w = \SI{6}{\micro m}$) oriented along the sample's [100] crystallographic direction were patterned using optical lithography and subsequent Ar ion etching. The device layout and connection scheme are depicted schematically in Fig. \ref{fig:two}(a). As a reference, we first investigate the SMR in the pure spin-current
system CoO($\SI{5}{nm}$)/Pt(\SI{2}{nm}) using angular-dependent MR scans at $\SI{200}{K}$ with an applied magnetic field of $\mu_0{H}=\SI{11}{T}$, which is sufficient to overcome the spin-flop field of the CoO (Fig. \ref{fig:two})~\cite{baldrati2020efficient}. We rotate the magnetic field within the sample plane and vary the angle $\alpha$ between the applied charge current and the magnetic field, as defined in Fig. \ref{fig:two}(a). The direction of the Néel vector $\boldsymbol{n}$ can be probed electrically via the transverse MR signal, which is proportional to the in-plane components $n_x$ and $n_y$~\cite{baldrati2018full,hoogeboom2017negative,Fischer2018}. The measured electrical signal is in line with previous measurements on CoO/Pt, with two distinct contributions: i) a hysteretic component around the directions of the magnetic hard axes of the sample, correlated with a negative SMR signal (Fig. \ref{fig:two}(b))~\cite{baldrati2020efficient, grzybowski2023antiferromagnetic}. This feature confirms the fourfold in-plane magnetic anisotropy of the CoO layer, induced by substrate strain, with two magnetic easy axes in the (001) plane ([110] and [$\bar{1}10$]) and an out-of-plane hard axis along the [001] direction; and (ii) a non-hysteretic sinusoidal background signal, correlated with a positive SMR, the origin of which according to previous investigations is not directly correlated with the Néel vector orientation in CoO~\cite{Schmitt2026}.

Inserting a Cu interlayer between the AFM CoO and spin-current source Pt systematically modifies both the amplitude and functional form of the MR signals as a function of Cu thickness $t_{\mathrm{Cu}}$. For the transverse component (Figs.~\ref{fig:two}(c) and (d)), the sinusoidal signal reverses sign as $t_{\mathrm{Cu}}$ increases: at $t_{\mathrm{Cu}}=\SI{0}{nm}$, $R_{xy}$ exhibits a maximum when $\boldsymbol{H} \parallel [110]$ ($\alpha = \SI{45}{\degree}$); at $t_{\mathrm{Cu}}=\SI{3}{nm}$, the sinusoidal component nearly vanishes, producing a flat response around $\alpha = \SI{45}{\degree}$; and at $t_{\mathrm{Cu}}=\SI{6}{nm}$, a minimum replaces the initial maximum at the same angle. 

Systematic measurements of $R_{xy}$ across intermediate Cu thicknesses (see supplementary information-2) reveal a continuous evolution: the positive sinusoidal amplitude decreases monotonically with increasing $t_{\mathrm{Cu}}$, vanishes near $t_{\mathrm{Cu}} \approx \SI{3}{nm}$, and transitions to a progressively stronger negative amplitude for $t_{\mathrm{Cu}} > \SI{3}{nm}$.

Considering the hysteretic signal that is clearly related to the orientation  of the Néel vector, we find that it becomes more pronounced upon insertion of a Cu layer in the transverse MR geometry. Moreover, the sign of the hysteretic component reverses with the introduction of a thin Cu interlayer,
and the amplitude of this inverted hysteretic signal further increases with increasing Cu thickness. The evolution of the signal can be understood as the superposition of two MR contributions arising from two distinct mechanisms. In our recent work on $\mathrm{CoO/Cu^*}$~\cite{Schmitt2026}, we have demonstrated that dynamic OAM i.e., orbital current, can effectively couple to the orbital-momentum-dominated magnetization in CoO, resulting in a pronounced orbital magnetoresistance (OMR) signal. Comparing this reported OMR to the transverse MR signal observed here (Fig. \ref{fig:two}(d)), we find that the CoO($\SI{5}{nm}$)/Cu($\SI{6}{nm}$)/Pt($\SI{2}{nm}$) sample exhibits a signal whose sinusoidal and hysteretic components, specifically their amplitude ratio, closely match those of the pure OMR observed in CoO($\SI{5}{nm}$)/$\mathrm{Cu(\SI{6}{nm})}$ as we reported in Ref.~\cite{Schmitt2026}. The absolute magnitude of the resistance change is comparable in both cases; however, the longitudinal resistance in the sample containing an additional Pt layer is slightly lower, resulting in a comparably larger relative change compared to pure CoO/Cu*. Notably, the hysteretic component associated with the spin-flop transition remains consistent across both systems, indicating that the underlying physical mechanisms are similar to those identified in our previous studies~\cite{Schmitt2026}, namely the generation of large OAM currents at Cu-oxide interface and their efficient interaction with orbital magnets.

The similarity of the MR effects observed in the two samples indicates that, in addition to the spin-based SMR signal and the spin current generated in the heavy metal Pt, an orbital current that is related to the Cu interlayer on the CoO and consequently an OMR contribution is also present. In contrast to previous work, this is here shown even in the absence of intended Cu oxidation. To interpret the observed results, the Pt-related spin contribution must be distinguished from the Cu-related orbital contribution. By varying the Cu thickness while keeping Pt fixed, the current distribution between the two metallic layers, and thus the relative contributions of SMR and OMR, can be tuned. Thus, the overall MR response, as a combination of these two contributions, evolves according to the interplay of the effects generated by the currents flowing in the two Cu and Pt layers.

\begin{figure*}
 \includegraphics[width=8cm]{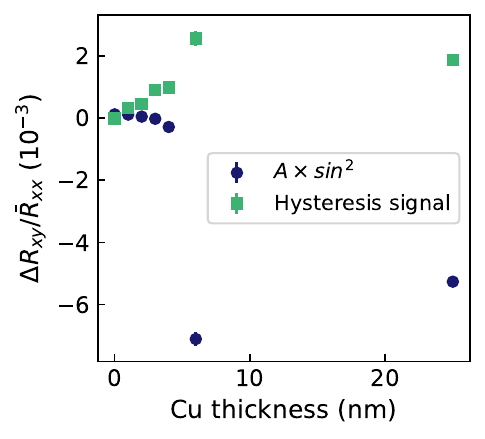}
 \caption{Amplitudes of the hysteretic (green squares) and sinusoidal (blue discs) signal components as a function of Cu interlayer thickness.}
 \label{fig:three}
\end{figure*}

Since the OMR in CoO/Cu* was found to be significantly stronger than the SMR in CoO/Pt, we extracted the amplitudes of both the hysteretic and sinusoidal components as a function of Cu interlayer thickness from the transverse $R_{xy}$ measurements. This analysis allows us to check whether the amplitude increases with the amount of current flowing through the Cu layer, as would be expected from an enhanced contribution of OMR. The results, shown in Fig. \ref{fig:three}, reveal that both signal amplitudes increase with Cu thickness up to $\SI{6}{nm}$. This trend is in agreement with the hypothesis that the overall MR signal strengthens due to enhanced coupling of OAM to the orbital moments in CoO as a relatively larger fraction of the current is flowing through the Cu layer. Notably, even though the amplitude of the sinusoidal signal increases for large $t_{\mathrm{Cu}}$, it approaches zero for interlayer thicknesses of $2 - \SI{3}{nm}$, demonstrating that one can eliminate the component unrelated to the Néel vector orientation when choosing an appropriate system geometry, keeping only the signal directly related to the Néel vector orientation in CoO.

To quantify the behaviour of the signals, we use the Fuchs-Sondheimer model to estimate size-dependent resistivities and sheet resistances~\cite{fuchs1938conductivity, sondheimer2001mean} and calculate the expected current distribution between the Cu and Pt layers (see supplementary material-3). We find that for a $\SI{3}{nm}$ thick Cu layer, the sheet resistance is $\SI{33}{\Omega/\mathrm{square}}$, resulting in roughly $82\%$ of the total current flowing through Cu. Increasing the Cu thickness to $\SI{6}{nm}$ reduces its sheet resistance to $\SI{10}{\Omega/\mathrm{square}}$, increasing the current fraction flowing in Cu to about $94\%$.

The saturation of the signal amplitudes observed at a Cu thickness of $\SI{6}{nm}$, together with the absence of further enhancement at larger Cu interlayer thicknesses, indicates that beyond this point the vast majority of the current already flows through the Cu interlayer. This behavior is consistent with the higher intrinsic conductivity of Cu compared to Pt and supports the quantitative estimates.

We can, thus, conclude that the measured signal is dominated by the OMR, consistent with the MR response in Fig. \ref{fig:two}(d), which closely resembles that observed in pure $\mathrm{CoO/Cu^*}$ without any Pt~\cite{Schmitt2026}. A large orbital current, however, is not expected to be generated in pure Cu~\cite{Shilei2020}, but only from some interaction of Cu with oxygen~\cite{go2021orbital}. In the present case, the Cu interlayer is capped by a continuous $\SI{2}{nm}$ Pt film, which effectively prevents oxidation from the environment. However, a sizable orbital current is still observed. We find no significant interface oxidation in the CoO/Cu/Pt trilayers investigated here (see Supplementary Information-4). Given that the Cu is capped well with Pt, one also does not expect any surface oxidation of the Cu. Looking at the Cu/Pt interface, we do not see the formation of a significant CuO layer \cite{ding2025generation,Schmitt2026}. But given the proximity of Cu and CoO, our results indicate that the formation of O–Cu bonds at this interface is sufficient to generate a large orbital current via the OREE. This conclusion is in line with theoretical predictions that an ultrathin oxygen layer at the CoO/Cu interface is sufficient to induce a colossal OREE~\cite{go2021orbital}.
These findings thus demonstrate that the CoO/Cu interface alone is sufficient for generating large orbital currents and producing a giant OMR, without relying on natural and uncontrolled surface oxidation. This opens the way to well-defined, stable, cheap and environmentally friendly orbitronic devices, as the capping Pt layer could be substituted by any suitable, inexpensive material that prevents additional surface oxidation of the Cu.

\section{\label{sec:conclusion}Conclusion}
In conclusion, we find that both orbital and spin Hall MR, contribute to the measured MR signal in CoO/Cu/Pt trilayer heterostructures. Our TEM imaging shows the formation of epitaxial growth of CoO/Cu on MgO with well-defined interfaces. Angular-dependent MR measurements reveal a systematic evolution of the transverse signal with Cu interlayer thickness. By varying the Cu thickness while keeping Pt fixed, the relative amount of charge current contributing to the generation of orbital and spin current varies respectively, thus changing the resulting MR signal. Since the Cu surface is fully capped by a continuous Pt layer and no  oxidation of Cu is detected at Cu/Pt interface,  the generation of dynamic OAM can be attributed solely to the OREE effect at the CoO/Cu interface. Our experimental results point to the formation of chemical bonds of the Cu atoms located directly at the interface with $\mathrm{O}^{2-}$-ions from the CoO. The generation of the larger orbital signals can thus be traced to this formation of an interfacial Cu-O chemical bonding. These findings highlight the potential of $\mathrm{CoO/Cu^*/\mathrm{capping\ layer}}$ devices for next-generation orbitronic applications.

\section{Author contributions}
The CoO samples were grown by CS with the assistance of TKi supervised by ES. CS and SK fabricated the devices and performed the electrical measurements. EG assisted with the experiments. CS analyzed the data with inputs from HM, EG, SK, OK, TKu and MK. CS and MK wrote the manuscript. MK proposed and supervised the whole project. All authors commented on the manuscript.

\section{acknowledgements}
The authors thank the DFG (Spin+X (A01, B02) TRR 173-268565370 and Project No. 358671374), the Horizon 2020 Framework Programme of the European Commission under FETOpen Grant Agreement No. 863155 (s-Nebula); the European Research Council Grant Agreement No. 856538 (3D MAGiC); and the Research Council of Norway through its Centers of Excellence funding scheme, Project No. 262633 ”QuSpin”. The study has also been supported by the European Horizon Europe Framework Programme under an EC Grant Agreement N°101129641 ”OBELIX”. This research is also a part of the ORBIS project, which is funded by the European Union's Horizon under the Marie Sk\l odowska-Curie Actions (MSCA), Grant Agreement No. 101226840. The work at the University of Tokyo was supported by CREST (Nos. JPMJCR20C1 and JPMJCR20T2) from JST, Japan, Grant-in-Aid for Scientific Research (S) (No. JP19H05600), Grant-in-Aid for Transformative Research Areas (No. JP22H05114) from JSPS KAKENHI, Japan, Grant-in-Aid for Scientific Research (B) (No. JP24K01326) from JSPS KAKENHI, Japan, Institute for AI and Beyond of the University of Tokyo, and Sumitomo Chemical.

\bibliography{main}
\clearpage
\setcounter{figure}{0}
\renewcommand{\thefigure}{S\arabic{figure}}

\newpage

{\Huge Supplementary Information}

\section{S.I. 1 Sample preparation and measurement procedure}

The CoO thin films were deposited on MgO(001) substrates using a fully automated ULVAC QAM 4 sputtering system operated at a base pressure below $1.2 \times \SI{E-5}{Pa}$. Prior to deposition, the MgO substrates were annealed at $\SI{770}{\degree C}$ for 2 h. After cooling to the deposition temperature, CoO was grown by reactive magnetron sputtering from a Co target in an Ar ($\SI{15}{sccm}$) and $\mathrm{O_2}$ ($\SI{2}{sccm}$) atmosphere at $\SI{573}{\degree C}$ using an RF power of $\SI{150}{W}$. After completion of the CoO deposition and cooling to room temperature, a Cu interlayer of the desired thickness and a $\SI{2}{nm}$ Pt capping layer were deposited in-situ, without breaking vacuum. These layers were prepared at room temperature in a separate chamber by DC magnetron sputtering.\\
Prior to patterning, the as-grown samples were cleaned with acetone and isopropanol. The device structures were defined by photolithography using a DMO MicroWriter ML3 and a negative photoresist, followed by Ar$^+$ ion milling. The resulting Hall bar geometry consists of a $\SI{6}{\micro m}$ wide main bar and two $\SI{2}{\micro m}$ wide transverse arms, separated by $\SI{6}{\micro m}$ and aligned along one of the magnetic hard axes of the sample.\\
For electrical characterization, the sample was mounted on a piezoelectric rotator inside a variable temperature insert (VTI) placed in a superconducting magnet providing magnetic fields up to $\SI{15}{T}$. Angular-dependent measurements were performed by rotating a constant magnetic field of $\SI{11}{T}$ within the sample plane. A charge current of $\SI{105}{\micro A}$ was applied to the $\SI{6}{\micro m}$ wide bar using a Keithley 6221 current source, alternating between positive and negative polarity. The longitudinal and transverse voltages were measured in the two transverse arms with a Keithley 2182A nanovoltmeter.\\
Each reported voltage represents the average of 100 measurements, each being the mean of signals measured with positive and negative current polarity. The error bars denote the standard deviation of these measurements. The applied current was modulated as a square-wave function of time. A delay was introduced between current reversal and data acquisition to suppress artifacts arising from the current source rise time and capacitive effects of the Hall structure. By evaluating the difference between the voltages measured for opposite current polarities, thermally induced contributions, such as the spin Seebeck effect, were effectively eliminated. The resulting voltage variation was analyzed as a function of the angle between the charge current and the magnetic field.

\section{S.I. 2 Additional SMR measurements}

The transverse $R_{xy}$ measurements were performed on 7 samples with Cu thicknesses varying between $\SI{0}{nm}$ and $\SI{6}{nm}$ in $\SI{1}{nm}$ steps and on an additional sample with $\SI{25}{nm}$ Cu interlayer to check for saturation of the signal as a function of Cu thickness. We find that the sinusoidal signal continuously transitions from a maximum at 45° to a minimum over the range of $\SI{0}{nm}$ Cu to $\SI{6}{nm}$ Cu. The sign of the signal change in the hysteresis curve also reverses and becomes more pronounced with increasing Cu layer thickness. The results on the two samples with $d = \SI{6}{nm}$ and $\SI{25}{nm}$ Cu interlayer show a very similar behavior regarding the amplitude and functional dependence of the MR signal.

\begin{figure}[h!]
 \includegraphics[width=16cm]{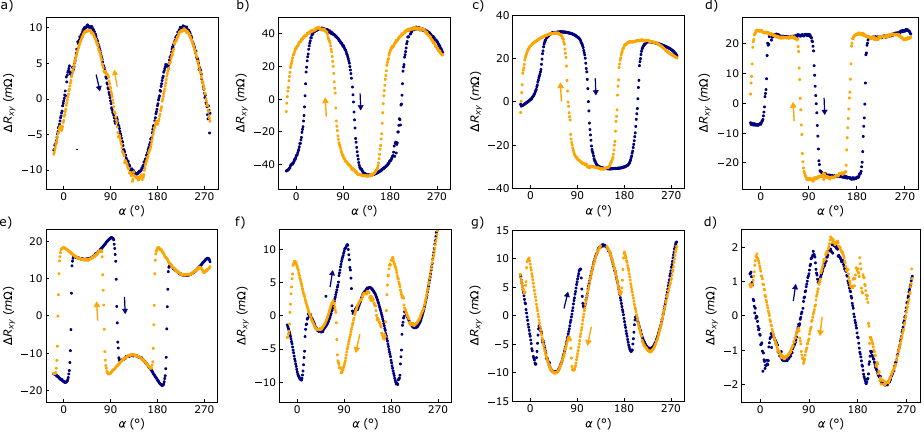}
 \caption{Angular dependent transverse magnetoresistance measurements for several samples with different Cu interlayer thickness. (a) $\SI{0}{nm}$, (b) $\SI{1}{nm}$, (c) $\SI{2}{nm}$, (d) $\SI{3}{nm}$, (e) $\SI{4}{nm}$, (f) $\SI{5}{nm}$, (g) $\SI{6}{nm}$, (h) $\SI{25}{nm}$.}
 \label{fig:SMR}
\end{figure}

\section{S.I. 3 Current flow through the Cu and Pt layers}

To estimate the amount of current flowing through the Cu and the Pt, respectively and to consequently estimate the contributions from spin and orbital current to the measured MR signal, we estimated the resistivity of the individual thin-film layers according to the Fuchs-Sondheimer model~\cite{fuchs1938conductivity, sondheimer2001mean}, which describes the thickness-dependent resistivity in thin metal films, accounting for surface scattering of conduction electrons. Here, we employ a simplified linear approximation, which, for ultrathin films with thicknesses smaller than the electron mean free path, tends to underestimate the resistance increase. Consequently, the current fraction in Cu is likely slightly underestimated, whereas the current fraction in Pt is likely slightly overestimated.

For these calculations, the bulk resistivities of Cu and Pt was considered as $\rho_{\mathrm{Cu,bulk}} = 1.68 \times 10^{-8} \ \Omega\mathrm{m}$ and $\rho_{\mathrm{Pt,bulk}} = 1.06 \times 10^{-7} \ \Omega\mathrm{m}$, respectively~\cite{lide2004crc}. The electron mean free path $\ell$ in Cu is $\ell_{\mathrm{Cu}} = 39\ \mathrm{nm}$~\cite{gall2016electron} and in Pt $\ell_{\mathrm{Pt}} = 10 \ \mathrm{nm}$~\cite{dushenko2018tunable}.

For a metal film of thickness $d$, with bulk resistivity $\rho_0$, and electron mean free path $\ell$, the simplified Fuchs–Sondheimer expression for the increased resistivity $\rho(d)$ is:

$$
\rho(d) = \rho_0 \left( 1+ \frac{3\ell}{8d}\right).
$$

This equation assumes fully diffuse surface scattering.

Calculating this for $\SI{3}{nm}$ and $\SI{6}{nm}$ Cu interlayers, as well as for the $\SI{2}{nm}$ Pt layer yields:

$$
\rho_{\mathrm{Cu}}(\SI{6}{nm}) = \SI{5.8e{-8}}{\Omega m},
$$
$$
\rho_{\mathrm{Cu}}(\SI{3}{nm}) = \SI{9.9e{-8}}{\Omega m},
$$
$$
\rho_{\mathrm{Pt}}(\SI{2}{nm}) = \SI{3.0e{-7}}{\Omega m}.
$$

For comparison, we determined the resistivities experimentally in bilayers comprising $\SI{5}{nm}$ CoO and either $\SI{6}{nm}$ Cu, $\SI{3}{nm}$ Cu, or $\SI{2}{nm}$ Pt:

$$
\rho_{\mathrm{Cu, Exp}}(\SI{6}{nm}) = \SI{4.9e{-8}}{\Omega m},
$$
$$
\rho_{\mathrm{Cu, Exp}}(\SI{3}{nm}) = \SI{50.4e{-8}}{\Omega m},
$$
$$
\rho_{\mathrm{Pt, Exp}}(\SI{2}{nm}) = \SI{23.5e{-8}}{\Omega m}.
$$
\noindent
These values, however, cannot be directly compared to the CoO/Cu/Pt trilayers studied here, since in the calibration samples the Cu layer was uncapped and therefore subject to oxidation. The resistivities of Pt and Cu(6) remain reasonably close to the theoretical predictions (within $\sim$20\%), but the Cu(3) film exhibits a drastically increased resistivity due to the oxidation of a significant part of the ultrathin layer. Because of this effect we rely on the theoretically calculated resistivities when determining the current distribution in the CoO/Cu/Pt structures, which appear to be a good approximation.

From the resistivities, we calculated with the Fuchs-Sondheimer model the sheet resistance of each layer can be obtained as $R_{\mathrm{s}} = \rho / t$. The current distribution between two parallel conducting layers is inversely proportional to their respective sheet resistances, such that:

$$
I_{\mathrm{Cu}} = I_{\mathrm{total}} \times \frac{1/R_{\mathrm{s,Cu}}}{1/R_{\mathrm{s,Cu}} + 1/R_{\mathrm{s,Pt}}} = I_{\mathrm{total}} \times \frac{R_{\mathrm{s,Pt}}}{R_{\mathrm{s,Cu}} + R_{\mathrm{s,Pt}}}.
$$

Using this relation, we find that in the sample with Cu($\SI{3}{nm}$)/Pt($\SI{2}{nm}$), approximately 82\% of the total current flows through the Cu layer. For Cu($\SI{6}{nm}$)/Pt($\SI{2}{nm}$), this fraction increases to about 94\%. These values likely represent lower estimates, since the Cu and Pt layers differ in thickness by a factor of three, and the simplified linear form of the Fuchs-Sondheimer equation tends to underestimate the resistance increase in Pt more strongly than in Cu.

\section{S.I. 4 Characterization of interface oxidation}

\begin{figure*}
 \includegraphics[width=15cm]{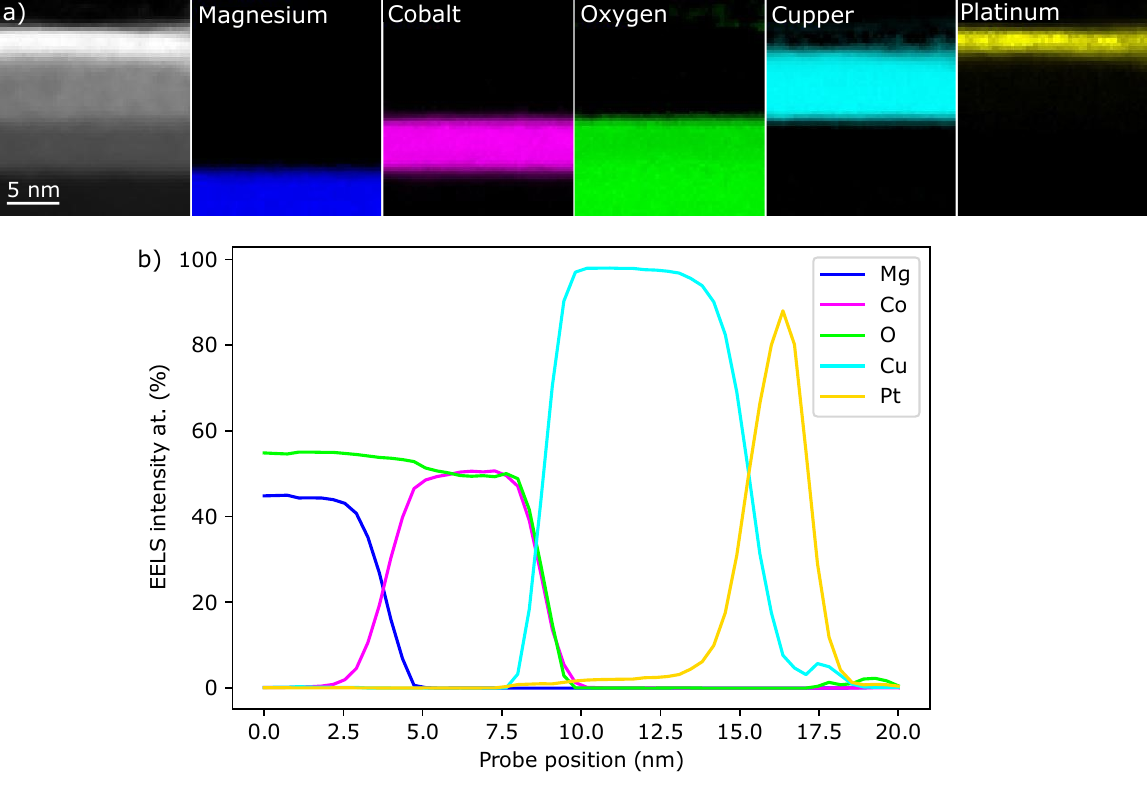}
 \caption{(a) HAADF-STEM image and corresponding elemental EELS maps of Mg (blue), Co
(pink), O (green), Cu (turquoise) and Pt (yellow). (b) Corresponding atomic fraction profiles along the growth direction.}
 \label{fig:STEM}
\end{figure*}

To determine a possible oxidation at the CoO/Cu interface, scanning transmission electron microscopy (STEM) with a high-angle annular dark-field detector (HAADF) and electron energy-loss spectroscopy (EELS) measurements were performed, as shown in Fig. \ref{fig:STEM}. A cross-sectional TEM lamella was prepared by Ga$^+$ focused ion beam (FIB) milling at $\SI{30}{kV}$ using a Thermo Fisher Scientific Helios dual-beam platform FIB-SEM. The final thinning step was performed at $\SI{5}{kV}$ to minimize surface damage. The HAADF-STEM and EELS measurements were conducted using a TFS Spectra TEM operated at $\SI{300}{kV}$ and equipped with a Gatan Continuum imaging filter (GIF). The EELS elemental maps and corresponding atomic fraction profiles extracted along the growth direction shown in Fig. \ref{fig:STEM} show that within the experimental resolution no significant oxidation of the Cu layer at the CoO/Cu interface can be observed. 

For the strain analysis presented in Fig. \ref{fig:one}(c) of the main text, we performed geometrical phase analysis (GPA). GPA is an image processing technique applied to high-resolution images, which uses Fourier transforms and apertures around Bragg spots to reconstruct the displacement and strain fields associated with different sets of lattice planes. In this work, GPA was performed using the Digital Micrograph software (DM gatan) with a freely available plugin (https://er-c.org/index.php/software/stem-data-analysis/gpa/). 

\end{document}